\documentclass[aps,prl,reprint,groupedaddress,amsmath,amssymb,floatfix,footinbib]{revtex4-2}
\usepackage[T1]{fontenc}
\usepackage{lmodern}
\usepackage{microtype}
\usepackage{needspace}
\usepackage[hidelinks]{hyperref}
\hypersetup{pdftitle={Nonperturbative Ground-State Energy from Propagators for Any Interaction},pdfauthor={Alexander Kruchkov and F. Duncan M. Haldane}}

\begin{document}
\title{Nonperturbative Ground-State Energy from Propagators for Any Interaction}
\author{Alexander Kruchkov}
\author{F. Duncan M. Haldane}
\affiliation{Department of Physics, Princeton University, Princeton, New Jersey 08544, USA}
\date{\today}

\begin{abstract}
Fractional Chern insulators motivate the search for exact energy relations in strongly correlated flat bands with many-body interactions. Haussmann, Mori, and MacDonald obtained such a relation for \textit{pairwise} interactions. However, additional three-body and higher-body terms invalidate that energy expression. Nevertheless, three-body and higher-body interaction terms enter the physics of greatest interest in fractional Chern insulators and other competing quantum phases. We obtain a \textit{nonperturbative} relation between the interacting ground-state energy and the exact propagators for arbitrary many-body couplings beyond the pairwise case. \end{abstract}
\maketitle
\setlength{\abovedisplayskip}{5pt plus 3pt minus 2pt}
\setlength{\belowdisplayskip}{\abovedisplayskip}
\setlength{\abovedisplayshortskip}{\abovedisplayskip}
\setlength{\belowdisplayshortskip}{\abovedisplayskip}

Fractional Chern insulators are phases selected by interactions in partially filled topological bands. When the one-body Hamiltonian of a topological flat band is suppressed, the noninteracting many-body space is degenerate, and interactions select the ground state within that space. For pair interactions restricted to a spin-polarized Landau level, Haussmann, Mori, and MacDonald~\cite{Haussmann1996, MacDonald1999} gave the ground-state energy in terms of the exact spectral functions. Their \textit{nonperturbative} result \cite{Haussmann1996} can be viewed as the interacting flat-band case of the Galitskii--Migdal  energy-propagator relation \cite{GalitskiiMigdal1958,Koltun1972},

\begin{equation}
\frac{E_{\text{G.S.}}(\nu)}{N_\phi}=\frac{1}{2}\int\limits_{-\infty}^{\mu}\!\mathrm{d}\omega\,
\omega \rho(\omega),\qquad
\nu=\int\limits_{-\infty}^{\mu}\!\mathrm{d}\omega \,  \, \rho(\omega).
\label{eq:GM}
\end{equation}
The spectral density is specified by the imaginary part of the retarded propagator through $- \frac{1}{\pi
} \operatorname{Im}\operatorname{Tr}[\mathcal{G}^{\mathrm R} (\omega) ] \equiv N_\phi \, \rho(\omega) $, where $N_\phi$ orbitals contain $N$ particles at Landau-level filling $\nu=N/N_\phi$.
This simple identity is exact and nonperturbative, and thus applies to correlated flat-band phases, including fractional quantum Hall states.
Its scope is restricted to pairwise ($\mathcal V_2$) interactions. 

The extension of the ground-state energy--propagator relation in Eq.~\eqref{eq:GM} beyond pairwise interactions, to a Hamiltonian $\mathcal H = \mathcal H_0 + \mathcal V_2 + \mathcal V_3 + \mathcal V_4 + \cdots$, remains a fundamental unresolved field-theoretic problem. For a flat band ($\mathcal H_0=0$), each additional interaction rank contributes a distinct correction to Eq.~\eqref{eq:GM},
\begin{equation}
\begin{aligned}
E_{\text{G.S.}}& = \frac{1}{2}\int_{-\infty}^{\mu}\!\mathrm d\omega\,
\omega\operatorname{Tr}[\hat{\rho}(\omega)]
-\frac{1}{2}\langle\mathcal V_3\rangle-\langle\mathcal V_4\rangle\\
&-\frac{3}{2}\langle\mathcal V_5\rangle-2\langle\mathcal V_6\rangle
-\frac{5}{2}\langle\mathcal V_7\rangle-3\langle\mathcal V_8\rangle
-\frac{7}{2}\langle\mathcal V_9\rangle+ \cdots.
\end{aligned}
\label{eq:GMthreeCorrection}
\end{equation}
The different coefficients assigned to the higher-body terms $\mathcal V_n$ in \eqref{eq:GMthreeCorrection} prevent the unweighted sum
\begin{equation}
\begin{aligned}
E_{\text{G.S.}}= \langle\mathcal V_2\rangle
+ \langle\mathcal V_3\rangle+ \langle\mathcal V_4\rangle + \langle\mathcal V_5\rangle + \langle\mathcal V_6\rangle + \langle\mathcal V_7\rangle +\cdots
\end{aligned}
\label{eq:flatEnergySum}
\end{equation}
from being expressed \textit{nonperturbatively} by a ground-state (G.S.) energy--propagator relation of the form Eq.~\eqref{eq:GM} for generic mixtures of distinct ranks in terms of the zeroth and first moments of $\mathcal G (\omega)$.

Higher-body interactions of rank $n>2$ are, however, essential to both the stability of quantum phases and the character of their excitations.
In fractional Chern insulators, the lattice counterparts of fractional quantum Hall states, such terms arise naturally when the effective Hamiltonian accounts for processes beyond projection of the pair interaction. Virtual transitions through remote bands can generate effective three-body terms as well as corrections to the pair interaction~\cite{SodemannMacDonald2013,ZhangSong2024}. 
Parent Hamiltonians provide a second realization. Within a Landau level, Haldane pseudopotentials resolve the interaction into allowed two-particle relative-angular-momentum channels, each with its assigned energy~\cite{Haldane1983}. The same projector construction applies to multiparticle interactions~\cite{SimonRezayiCooper2007}. In the lowest Landau level, the positive sum of three-body projectors onto the smallest allowed fermionic relative angular momentum, $m=3$, admits the Moore--Read state~\cite{MooreRead1991} as a densest zero-energy state at the appropriate flux for even particle number~\cite{GreiterWenWilczek1992}. The Read--Rezayi sequence follows from related $(k+1)$-body parent Hamiltonians~\cite{ReadRezayi1999}.
Exact diagonalization of model Chern bands with three-body interactions also provides evidence for Moore--Read-like states~\cite{ZhangSong2024}.

In this Letter, we address the \textit{nonperturbative} extension of the ground-state energy--propagator relation and obtain an exact result for interactions of arbitrary many-body rank $n>2$ in full interacting Hamiltonian $\mathcal H = \mathcal H_0 + \sum_n \mathcal V_n$. The construction uses the evolution of the Green's function $\mathcal G_{\lambda}$ along an auxiliary interaction flow $\lambda$, expressed through either the Hamiltonian $\lambda$-path $\mathcal H_{\lambda}=\mathcal H_0+\sum_{n\ge2}\lambda^{n-1}\mathcal V_n$, or the field rescaling $(c^{\dag}_{\alpha},c^{}_{\alpha})\to\sqrt\lambda(c^{\dag}_{\alpha},c^{}_{\alpha})$, with the Hamiltonian and field normalizations specified below,
\begin{equation}
E_{\text{G.S.}}=\int_0^1\!\mathrm d\lambda
\int_{-\infty}^{+\infty}\!\frac{\mathrm d\omega}{2\pi i}\,
e^{i\omega0^+}\omega\operatorname{Tr}[\mathcal G_\lambda(\omega)].
\label{eq:mainintro}
\end{equation}
This identity gives the ground-state energy in terms of exact propagators without a perturbative expansion, assuming normal ordering, instantaneous many-body interactions, and the regularity conditions stated below.
The corresponding special cases recover Eqs.~\eqref{eq:GM} and \eqref{eq:GMthreeCorrection}. A smooth field  reparametrization leaves the ground state  energy unchanged.

\needspace{4\baselineskip} \textit{Formalism.---} Let canonical fermion operators obey $\{c^{}_{\alpha},c^{\dag}_{\beta}\}=\delta_{\alpha\beta}$ and $\{c^{}_{\alpha},c^{}_{\beta}\}=0$ in the orbital basis $\alpha, \beta$, and consider the many-body Hamiltonian
\begin{equation}
\mathcal{H}=\mathcal{H}_0+\sum_{n\ge2}\mathcal{V}_n,
\label{eq:Hfull}
\end{equation}
which is Hermitian and conserves total particle number. Each instantaneous, vacuum-normal-ordered term $\mathcal V_n$ contains $n$ creation and $n$ annihilation operators, with fixed interaction vertices. The one-body term has the form $\mathcal H_0=\sum_{\alpha,\beta}\mathcal H^{(0)}_{\alpha\beta}c^{\dag}_{\alpha} c^{}_{\beta}$, where $\alpha, \beta$ label orbitals.  We set $\hbar=1$ and denote the expectation of a Hermitian observable $\hat  {\mathcal X} $ by $\langle \hat{\mathcal X} \rangle$, evaluated in a normalized $N$-particle ground state.
Operator hats are omitted except where the operator identity need to be highlighted. 

The fermionic fields evolving in the Heisenberg picture, $c^{}_{\alpha}(t)=e^{i\mathcal Ht}c^{}_{\alpha} e^{-i\mathcal Ht}$, obey the equation of motion

\begin{align}
i\partial_t
c^{}_{\alpha}(t)=\,\left[c^{}_{\alpha}(t),\mathcal H\right].
\label{eq:HeisenbergEOM}
\end{align}
The same operator identity determines the time dependence of the propagator. With $\mathcal T$ denoting fermionic time ordering, the \textit{time-ordered} propagator and its Fourier transform in the orbital basis~\cite{Mahan} are defined by
\begin{align}
\mathcal{G}_{\alpha\beta}(t)&=-i\langle \mathcal T c^{}_{\alpha}(t) c^{\dag}_{\beta}(0)\rangle,
\label{eq:Gtime}\\
&=\int\limits_{-\infty}^{+\infty}\!\frac{\mathrm{d}\omega}{2\pi}
e^{-i\omega t}\mathcal{G}_{\alpha\beta}(\omega).
\label{eq:Gfourier}
\end{align}
The density matrix $\hat n$, with occupations $n_\alpha=n_{\alpha\alpha}$ and normalization $\operatorname{Tr}\hat n =N$, is obtained from the propagator by taking the negative-time limit $t \to 0^{-}$,
\begin{equation}
n_{\alpha\beta}\equiv\langle  c^{\dag}_{\beta} c^{}_{\alpha}\rangle=\int\limits_{-\infty}^{+\infty}\!\frac{\mathrm{d}\omega}{2\pi i}\,e^{i\omega0^{+}}\,\mathcal{G}_{\alpha\beta}(\omega).
\label{eq:occupation}
\end{equation}
The corresponding derivative limit $\lim _{t\to 0^{-}} \partial_t\mathcal{G}_{\alpha\beta}(t ) $ contains further \textit{nonperturbative} information. The propagator satisfies $\mathcal{G}_{\alpha\beta}(t)=i\langle
c^{\dag}_{\beta}c^{}_{\alpha}(t)\rangle$ for $t<0$. Equation~\eqref{eq:HeisenbergEOM} then gives the propagator equation of motion for $t<0$,
\begin{align}
i\partial_t\mathcal{G}_{\alpha\beta}(t < 0)&=i\langle  c^{\dag}_{\beta}\,\left[c^{}_{\alpha}(t),\mathcal{H}\right]\rangle,
\label{eq:negtEOM}\\
&=\int\limits_{-\infty}^{+\infty}\!\frac{\mathrm{d}\omega}{2\pi}
e^{-i\omega t}\,\omega  \, \mathcal{G}_{\alpha\beta}(\omega).
\label{eq:negtFourier}
\end{align}
The limit $\lim _{t\to 0^{-}} \partial_t\mathcal{G}_{\alpha\beta}(t ) $ defines the first-moment matrix
\begin{equation}
\mathcal W_{\alpha\beta}\equiv\langle  c^{\dag}_{\beta}\,\left[c^{}_{\alpha},\mathcal{H}\right]\rangle
=\int\limits_{-\infty}^{+\infty}\!\frac{\mathrm{d}\omega}{2\pi
i}\,e^{i\omega0^{+}}\,\omega  \, \mathcal{G}_{\alpha\beta}(\omega).
\label{eq:moment}
\end{equation}
Important observation is that each matrix element here has units of \textit{energy}.

Define a linear map $\mathcal{D}$ on Fock-space operators $\mathcal X$ by

\begin{equation}
\mathcal{D}\mathcal X\equiv\sum_{\alpha}c^{\dag}_{\alpha}\,\left[c^{}_{\alpha},\mathcal X\right].
\label{eq:Ddef}
\end{equation}
The ground-state expectation of $\mathcal{D} \mathcal H$ is

\begin{equation}
\langle\mathcal D\mathcal H\rangle
=\sum_\alpha\mathcal W_{\alpha\alpha}\equiv\mathcal W.
\label{eq:momentSum}
\end{equation}
Since $\mathcal H$ is Hermitian, Eq. \eqref{eq:momentSum} defines a real-valued scalar $\mathcal W$ with units of \textit{energy}. In an interacting state, this scalar generally differs from the ground-state energy. For the one-body Hamiltonian, the elementary commutator $\,\left[c^{}_{\alpha},c^{\dag}_{\gamma} c^{}_{\delta}\right]=\delta_{\alpha\gamma}c^{}_{\delta}$ gives
\begin{equation}
\,\left[c^{}_{\alpha},\mathcal{H}_0\right]=\sum_\beta\mathcal H^{(0)}_{\alpha\beta}c^{}_{\beta},\qquad
\mathcal{D}\mathcal{H}_0=\mathcal{H}_0.
\label{eq:kin}
\end{equation}
Thus $\mathcal W_0 \equiv \langle \mathcal{D}\mathcal{H}_0 \rangle = \langle \mathcal{H}_0 \rangle$ is the one-body energy evaluated in the chosen state.

\

\textit{Two-body inetractions.---} We first recover the ground-state energy relation \eqref{eq:GM} for a pairwise-interacting Hamiltonian. In terms of fully antisymmetrized matrix elements, the general pair interaction is
\begin{equation}
\mathcal{V}_2=\frac{1}{(2!)^2}
\sum_{\substack{\alpha_1,\alpha_2\\\beta_1,\beta_2}}
U^{\alpha_1\alpha_2}_{\beta_1\beta_2}\,
 c^{\dag}_{\alpha_1} c^{\dag}_{\alpha_2}c^{}_{\beta_2}c^{}_{\beta_1}.
\label{eq:V2}
\end{equation}
The tensor has antisymmetry within each index pair and Hermiticity under exchange of pairs; the unrestricted orbital sum has a factor $1/4$. For the monomial $\Omega^{\alpha_1\alpha_2}_{\beta_1\beta_2}=c^{\dag}_{\alpha_1}c^{\dag}_{\alpha_2}c^{}_{\beta_2}c^{}_{\beta_1}$, the commutator receives one contraction from each fermionic creator.
Left multiplication by $c^{\dag}_{\gamma}$, followed by summation, restores the original monomial in every contribution. The resulting identity is $\sum_\gamma c^{\dag}_{\gamma}\,\left[c^{}_{\gamma},\Omega^{\alpha_1\alpha_2}_{\beta_1\beta_2}\right]=2\Omega^{\alpha_1\alpha_2}_{\beta_1\beta_2}$. Reinstating the interaction tensor gives

\begin{align}
\mathcal{D}\mathcal{V}_2
&=\frac{1}{4}\sum_{\substack{\alpha_1,\alpha_2\\\beta_1,\beta_2}}
U^{\alpha_1\alpha_2}_{\beta_1\beta_2}\,
\sum_\gamma c^{\dag}_{\gamma}\,\left[c^{}_{\gamma},\Omega^{\alpha_1\alpha_2}_{\beta_1\beta_2}\right]\notag\\
&=\frac{1}{2}\sum_{\substack{\alpha_1,\alpha_2\\\beta_1,\beta_2}}
U^{\alpha_1\alpha_2}_{\beta_1\beta_2}\,\Omega^{\alpha_1\alpha_2}_{\beta_1\beta_2} \equiv 2\mathcal{V}_2.
\label{eq:Dtwo}
\end{align}
For the pairwise Hamiltonian $\mathcal{H}=\mathcal{H}_0+\mathcal{V}_2$, the energy characteristic $\mathcal W$ defined in \eqref{eq:momentSum} therefore becomes
\begin{equation}
\mathcal{W}=\langle \mathcal{H}_0\rangle+2\langle \mathcal{V}_2\rangle,
\label{eq:twocount}
\end{equation}
which counts the interaction expectation twice. The ground-state energy counts each Hamiltonian term only once,
\begin{equation}
E_{\text{G.S.}}=\langle \mathcal{H}_0\rangle+\langle \mathcal{V}_2\rangle.
\label{eq:twoEnergy}
\end{equation}
Elimination of $\langle\mathcal{V}_2\rangle$ between these expressions gives the Galitskii--Migdal  relation \cite{GalitskiiMigdal1958} for the ground-state energy of a pairwise-interacting system in terms of its exact propagator,

\begin{align}
E_{\text{G.S.}}&=\frac{1}{2}\bigl(\langle\mathcal H_0\rangle+\mathcal W\bigr)
=\frac{1}{2}\bigl(\operatorname{Tr}[\mathcal H_0\hat n]+\mathcal W\bigr),
\label{eq:GMmoments}\\
&=\frac{1}{2}\int\limits_{-\infty}^{+\infty}\!
\frac{\mathrm{d}\omega}{2\pi i}\,e^{i\omega0^+}\,
\operatorname{Tr}[(\omega\mathbf1+\mathcal H_0)\mathcal G(\omega)].
\label{eq:GMfull}
\end{align}
The propagator identity \eqref{eq:GMfull} underlies Haussmann-Mori-MacDonald formula \cite{Haussmann1996, MacDonald1999} for FQHE states ,  Eq \eqref{eq:GM}. Indeed, for a (strongly) interacting flat band, or several bands with a common energy, the derivation procedure sets $\mathcal{H}_0=0$. Next step, the time-ordered propagator can be written in the Lehmann representation
\begin{equation}
\mathcal G_{\alpha\beta}(\omega)
=\int\limits_{-\infty}^{\mu_-}\!\frac{\mathrm d\omega'\,\rho^-_{\alpha\beta}(\omega')}{\omega-\omega'-i0^+}
+\int\limits_{\mu_+}^{+\infty}\!\frac{\mathrm d\omega'\,\rho^+_{\alpha\beta}(\omega')}{\omega-\omega'+i0^+}.
\label{eq:specrep}
\end{equation}
Here  in the presence of a gap we have "lower" and "upper" chemical potentials,  $\mu_-=E_0(N)-E_0(N-1)$ and $\mu_+=E_0(N+1)-E_0(N)$.  Chemical potential in the gap means $\mu_-<\mu<\mu_+$, hence leading to the spectral separation  $\hat\rho^-=\Theta(\mu-\omega)\hat\rho (\omega) $ and $\hat\rho^+=\Theta(\omega - \mu)\hat\rho (\omega) $.  Substitution of Eq.~\eqref{eq:specrep} into Eq.~\eqref{eq:GMfull} in the flat-band limit gives

\begin{equation}
E_{\text{G.S.}}=\frac{1}{2}\!\int\limits_{-\infty}^{+\infty}\!
\frac{\mathrm d\omega}{2\pi i}e^{i\omega0^+}
\omega\operatorname{Tr}[\mathcal G(\omega)]
=\frac{1}{2}\!\int\limits_{-\infty}^{\mu_-}\!\mathrm d\omega\,
\omega\operatorname{Tr}[\hat{\rho}^-(\omega)].
\label{eq:flatPair}
\end{equation}
For a single Landau level, the orbital normalization $\operatorname{Tr}[\hat{\rho}(\omega)] =  N_\phi \rho(\omega) $ yields Eq.~\eqref{eq:GM}.

\

\textit{Three-body interactions.---} One might expect the ground-state energy formula \eqref{eq:flatPair} to survive the addition of interactions beyond the pairwise term.
The counting of operator contractions below shows why it does not.

Consider first a Hamiltonian whose only interaction is a three-body term,
\begin{equation}
\mathcal{H}=\mathcal{H}_0+\mathcal{V}_3.
\label{eq:Hthree}
\end{equation}
In vacuum normal order, the general three-body term has fully antisymmetrized matrix elements,

\begin{equation}
\mathcal V_3=\frac{1}{(3!)^2}
\sum_{\substack{\alpha_1,\alpha_2,\alpha_3\\\beta_1,\beta_2,\beta_3}}\!
U^{\alpha_1\alpha_2\alpha_3}_{\beta_1\beta_2\beta_3}\,
c^{\dag}_{\alpha_1}c^{\dag}_{\alpha_2}c^{\dag}_{\alpha_3}
c^{}_{\beta_3}c^{}_{\beta_2}c^{}_{\beta_1}.
\label{eq:V3}
\end{equation}
As in the pairwise case, the monomial $\Omega^{\alpha_1\alpha_2\alpha_3}_{\beta_1\beta_2\beta_3}$ denotes the operator product that enters the three-body interaction. Each contraction with a creator restores the monomial with a positive sign after left multiplication by $c^{\dag}_{\gamma}$ and summation. Hence $\mathcal D\Omega^{\alpha_1\alpha_2\alpha_3}_{\beta_1\beta_2\beta_3}
=3\Omega^{\alpha_1\alpha_2\alpha_3}_{\beta_1\beta_2\beta_3}$; reinstating the interaction tensor gives

\begin{align}
\mathcal{D}\mathcal{V}_3
&=\frac{1}{36}\!\sum_{\substack{\alpha_1,\alpha_2,\alpha_3\\\beta_1,\beta_2,\beta_3}}\!
U^{\alpha_1\alpha_2\alpha_3}_{\beta_1\beta_2\beta_3}\,\sum_\gamma c^{\dag}_{\gamma}\,\left[c^{}_{\gamma},\Omega^{\alpha_1\alpha_2\alpha_3}_{\beta_1\beta_2\beta_3}\right]\notag\\
&=\frac{1}{12}\sum_{\substack{\alpha_1,\alpha_2,\alpha_3\\\beta_1,\beta_2,\beta_3}}
U^{\alpha_1\alpha_2\alpha_3}_{\beta_1\beta_2\beta_3}\,
\Omega^{\alpha_1\alpha_2\alpha_3}_{\beta_1\beta_2\beta_3}\equiv 3\mathcal{V}_3.
\label{eq:Dthree}
\end{align}

\begin{table}[!t]
\caption{Single-rank energy relations [Eq.~\eqref{eq:nGM}]; each row uses its own ground-state propagator. }
\label{tab:ranks}
\setlength{\tabcolsep}{3pt}
\renewcommand{\arraystretch}{1.35}
\begin{ruledtabular}
\begin{tabular}{lc}
Hamiltonian & Ground-state energy $E_{\text{G.S.}}$\\
$\mathcal H=\mathcal H_0+\mathcal V_2$ & $\frac{1}{2}\!\int\!\frac{\mathrm{d}\omega}{2\pi i}\,\operatorname{Tr}{(\omega+\mathcal H_0)\mathcal G(\omega)}$\\
$\mathcal H=\mathcal H_0+\mathcal V_3$ & $\frac{1}{3}\!\int\!\frac{\mathrm{d}\omega}{2\pi i}\,\operatorname{Tr}{(\omega+2\mathcal H_0)\mathcal G(\omega)}$\\
$\mathcal H=\mathcal H_0+\mathcal V_4$ & $\frac{1}{4}\!\int\!\frac{\mathrm{d}\omega}{2\pi i}\,\operatorname{Tr}{(\omega+3\mathcal H_0)\mathcal G(\omega)}$\\
$\mathcal H=\mathcal H_0+\mathcal V_5$ & $\frac{1}{5}\!\int\!\frac{\mathrm{d}\omega}{2\pi i}\,\operatorname{Tr}{(\omega+4\mathcal H_0)\mathcal G(\omega)}$
\end{tabular}
\end{ruledtabular}
\end{table}
For the three-body Hamiltonian $\mathcal{H}=\mathcal{H}_0+\mathcal{V}_3$, the energy characteristic $\mathcal W$ defined in \eqref{eq:momentSum} is therefore
\begin{equation}
\mathcal{W}=\langle \mathcal{H}_0\rangle+3\langle \mathcal{V}_3\rangle,
\label{eq:threecount}
\end{equation}
while the ground-state energy is the unweighted sum
\begin{equation}
E_{\text{G.S.}}=\langle \mathcal{H}_0\rangle+\langle \mathcal{V}_3\rangle.
\label{eq:threeEnergy}
\end{equation}
Elimination of the interaction expectation $\langle \mathcal V_3 \rangle $, with Eq.~\eqref{eq:occupation} for the density matrix, gives
\begin{align}
E_{\text{G.S.}}&=\frac{1}{3}\bigl(2\langle\mathcal H_0\rangle+\mathcal W\bigr)
=\frac{1}{3}\bigl(2\operatorname{Tr}[\mathcal H_0\hat n]+\mathcal W\bigr),
\label{eq:threeMoments}\\
&=\frac{1}{3}\int\limits_{-\infty}^{+\infty}\!
\frac{\mathrm{d}\omega}{2\pi i}\,e^{i\omega0^+}\,
\operatorname{Tr}[(\omega\mathbf1+2\mathcal H_0)\mathcal G(\omega)].
\label{eq:threeGM}
\end{align}
This structure is similar to the  pairwise interaction identity, Eq.~\eqref{eq:GMfull}, yet with modified coefficients.

The correspondence between the relation \eqref{eq:threeGM} for the engineered Hamiltonian $\mathcal H = \mathcal H_0 + \mathcal V_3$ and the pairwise result raises the question of whether the same construction extends to the full Hamiltonian $\mathcal H = \mathcal H_0 + \mathcal V_2+ \mathcal V_3$. For a generic mixture, however, no universal linear formula of this form exists.

The obstruction is the distinct counting factors attached to $\mathcal V_2$ and $\mathcal V_3$ in a mixture. For a Hamiltonian with both two-body and three-body interactions,
\begin{equation}
\mathcal{H}=\mathcal{H}_0+\mathcal{V}_2+\mathcal{V}_3,
\label{eq:Hmixed}
\end{equation}
the energy characteristic $\mathcal W$ is fixed by Eqs.~\eqref{eq:kin}, \eqref{eq:Dtwo}, and \eqref{eq:Dthree} as
\begin{equation}
\mathcal{W}=\langle \mathcal{H}_0\rangle+2\langle \mathcal{V}_2\rangle+3\langle \mathcal{V}_3\rangle,
\label{eq:mixedCount}
\end{equation}
while the ground-state energy is
\begin{equation}
E_{\text{G.S.}}=\langle \mathcal{H}_0\rangle+\langle \mathcal{V}_2\rangle+\langle \mathcal{V}_3\rangle.
\label{eq:mixedEnergy}
\end{equation}
Subtracting Eq.~\eqref{eq:mixedCount} from twice the energy in Eq.~\eqref{eq:mixedEnergy} gives $2E_{\text{G.S.}}-\mathcal W
=\langle\mathcal H_0\rangle-\langle\mathcal V_3\rangle$. The pairwise expression must therefore be supplemented by the correction~\cite{Carbone2013}
\begin{equation}
\begin{aligned}
E_{\text{G.S.}}&=\frac{1}{2}\bigl(\langle\mathcal H_0\rangle+\mathcal W\bigr)
-\frac{1}{2}\langle\mathcal V_3\rangle\\
&=\frac{1}{2}\int\limits_{-\infty}^{+\infty}\!\frac{\mathrm d\omega}{2\pi i}\,
e^{i\omega0^+}\operatorname{Tr}[(\omega\mathbf1+\mathcal H_0)\mathcal G(\omega)]
-\frac{1}{2}\langle\mathcal V_3\rangle.
\end{aligned}
\label{eq:mixedGMCorrection}
\end{equation}
For generic many-body interactions, including both two-body and three-body terms,  the different weights of the two-particle and three-particle terms in $\mathcal W$ exclude a universal linear energy expression in these moments alone.

\

\textit{A class of many-body Hamiltonians with an exact ground-state energy--propagator relation (Table~\ref{tab:ranks}).} The structure of Eq.~\eqref{eq:threeGM} motivates asking which class of Hamiltonians with a specified interaction rank admits an exact relation of the type in Eq.~\eqref{eq:GM}. Take a many-body Hamiltonian containing a single interaction rank,
\begin{equation}
\mathcal{H}=\mathcal{H}_0+\mathcal{V}_n.
\label{eq:Hn}
\end{equation}
The interaction of this rank has the general form
\begin{equation}
\mathcal{V}_n=\frac{1}{(n!)^2}
\sum_{\substack{\alpha_1,\ldots,\alpha_n\\\beta_1,\ldots,\beta_n}}\!
U^{\alpha_1\cdots\alpha_n}_{\beta_1\cdots\beta_n}\,
c^{\dag}_{\alpha_1}\cdots c^{\dag}_{\alpha_n}
c^{}_{\beta_n}\cdots c^{}_{\beta_1}.
\label{eq:Vn}
\end{equation}
The tensor $U^{\alpha_1\cdots\alpha_n}_{\beta_1\cdots\beta_n}$ is antisymmetric in each index set, and the operator $\mathcal V_n$ vanishes in every sector with fewer than $n$ particles. For the operator monomial $\Omega^{\alpha_1\cdots\alpha_n}_{\beta_1\cdots\beta_n}
=c^{\dag}_{\alpha_1}\cdots c^{\dag}_{\alpha_n}
c^{}_{\beta_n}\cdots c^{}_{\beta_1}$, cancellation of the uncontracted contribution follows from $2n$ fermionic interchanges.
The contraction with the $j$th creator and its restoration by left multiplication each supply a sign $(-1)^{j-1}$. It follows that $\mathcal D\Omega^{\alpha_1\cdots\alpha_n}_{\beta_1\cdots\beta_n}
=n\Omega^{\alpha_1\cdots\alpha_n}_{\beta_1\cdots\beta_n}$, and therefore

\begin{align}
\mathcal{D}\mathcal{V}_n
&=\frac{1}{(n!)^2}\sum_{\substack{\alpha_1,\ldots,\alpha_n\\\beta_1,\ldots,\beta_n}}\!
U^{\alpha_1\cdots\alpha_n}_{\beta_1\cdots\beta_n}\,
\sum_\gamma c^{\dag}_{\gamma}\,\left[c^{}_{\gamma},\Omega^{\alpha_1\cdots\alpha_n}_{\beta_1\cdots\beta_n}\right]\notag\\
&=\frac{1}{n!(n-1)!}\sum_{\substack{\alpha_1,\ldots,\alpha_n\\\beta_1,\ldots,\beta_n}}\!
U^{\alpha_1\cdots\alpha_n}_{\beta_1\cdots\beta_n}\,
\Omega^{\alpha_1\cdots\alpha_n}_{\beta_1\cdots\beta_n}
\equiv n\mathcal{V}_n.
\label{eq:Dn}
\end{align}
The energy characteristic \eqref{eq:momentSum} of the interacting Hamiltonian $\mathcal{H}=\mathcal{H}_0+\mathcal{V}_n$ is thus
\begin{equation}
\mathcal{W}=\langle \mathcal{H}_0\rangle+n\langle \mathcal{V}_n\rangle,
\label{eq:nCount}
\end{equation}
while the corresponding ground-state energy is
\begin{equation}
E_{\text{G.S.}}=\langle \mathcal{H}_0\rangle+\langle \mathcal{V}_n\rangle.
\label{eq:nEnergy}
\end{equation}
Eliminating the common interaction expectation $\langle \mathcal{V}_n\rangle$ gives the ground-state energy

\begin{equation}
\begin{aligned}
E_{\text{G.S.}}&=\frac{1}{n}\bigl((n-1)\langle\mathcal H_0\rangle+\mathcal W\bigr)\\
&=\frac{1}{n}\int\limits_{-\infty}^{+\infty}\!\frac{\mathrm d\omega}{2\pi i}\,
e^{i\omega0^+}\operatorname{Tr}[(\omega\mathbf1+(n-1)\mathcal H_0)\mathcal G(\omega)].
\end{aligned}
\label{eq:nGM}
\end{equation}
This relation has the form of Eq.~\eqref{eq:GM}, extending the Galitsky--Migdal identity to a many-body interaction of any single rank $n$.
We also note that operator $\mathcal D $ acts as a rank-counting operator.

Formulas for the relation between ground state energy and their exact propagators for the four lowest interaction ranks are listed in Table~\ref{tab:ranks}. Each identity is exact for its stated Hamiltonian. Every additional interaction rank has different counting weight.

\

\textit{Nonperturbative ground-state energy--propagator relation for any interaction.---} The exact, nonperturnative identities for a single-rank $n>2$ manybody interaction Eq. \eqref{eq:nGM} leaves open its extension to a Hamiltonian with arbitrary interaction terms.
To address this,  we write the physical Hamiltonian as

\begin{equation}
\mathcal H=\mathcal H_0+\mathcal V_2+\mathcal V_3+\cdots
=\mathcal H_0+\sum_{n\ge2}\mathcal V_n.
\label{eq:HI}
\end{equation}
We now introduce an auxiliary Hamiltonian whose interaction coefficients vary smoothly over $0<\lambda\le1$, with the one-body term held fixed,

\begin{equation}
\mathcal H_\lambda=\mathcal H_0+\lambda\mathcal V_2+\lambda^2\mathcal V_3+\cdots
=\mathcal H_0+\sum_{n\ge2}\lambda^{n-1}\mathcal V_n.
\label{eq:Hlambda}
\end{equation}
As a consequence, the full Hamiltonian $\mathcal H$ of Eq.~\eqref{eq:HI} is recovered at $\lambda=1$. Differentiating Eq.~\eqref{eq:Hlambda}, then applying the counting identity Eq.~\eqref{eq:Dn}, gives the operator relation

\begin{equation}
\partial_\lambda[\lambda\mathcal H_\lambda]
=\mathcal H_0+\sum_{n\ge2}n\lambda^{n-1}\mathcal V_n
=\mathcal D\mathcal H_\lambda.
\label{eq:operatorDerivative}
\end{equation}
Since for physically-relevant systems at $\lambda =0$  quantity $\lambda\mathcal H_\lambda$ vanishes, one allows the operator identity~\eqref{eq:operatorDerivative} to be integrated in the form 
\begin{equation}
\mathcal H=\int_0^1\!\mathrm d\lambda\,
\mathcal D\mathcal H_\lambda.
\label{eq:operatorIntegral}
\end{equation}
At each $\lambda$, take the expectation in the actual fixed-$N$ ground state and define $\mathcal W(\lambda)=\langle\mathcal D\mathcal H_\lambda\rangle_\lambda$. Normalization and the eigenvalue equation cancel the derivatives of the state, so the Hellmann--Feynman theorem~\cite{Hellmann1937,Feynman1939} gives the energy derivative without a perturbative treatment,

\begin{equation}
\frac{\mathrm dE_{\text{G.S.}}(\lambda)}{\mathrm d\lambda}
=\langle\Psi_0^\lambda|\partial_\lambda\mathcal H_\lambda|\Psi_0^\lambda\rangle.
\label{eq:HellmannFeynman}
\end{equation}
Taking the ground-state expectation of Eq.~\eqref{eq:operatorDerivative} and applying Eq.~\eqref{eq:HellmannFeynman} then gives, by the product rule,

\begin{equation}
\frac{\mathrm d[\lambda E_{\text{G.S.}}(\lambda)]}{\mathrm d\lambda}
=E_{\text{G.S.}}(\lambda)
+\lambda\frac{\mathrm dE_{\text{G.S.}}(\lambda)}{\mathrm d\lambda}
=\mathcal W(\lambda).
\label{eq:energyFlow}
\end{equation}
With $\rho^-_E[\widehat{\mathcal H}_\lambda]\equiv\operatorname{Tr}[\hat\rho^-_\lambda(E)]$ Eqs.~\eqref{eq:operatorDerivative} and \eqref{eq:energyFlow} identify this derivative with $\langle\Psi_0^\lambda|\widehat{\mathcal H}_\lambda+\lambda\partial_\lambda\widehat{\mathcal H}_\lambda|\Psi_0^\lambda\rangle$. The spectral identities in Eqs.~\eqref{eq:moment} and \eqref{eq:specrep} express this expectation as the first moment, giving

\begin{equation}
\int_{-\infty}^{\mu^-_\lambda}\!\mathrm dE\,E\,\rho^-_E[\widehat{\mathcal H}_\lambda]
 = \langle\Psi_0^\lambda|\partial_\lambda[\lambda\widehat{\mathcal H}_\lambda]|\Psi_0^\lambda\rangle.
\label{eq:spectralHellmannFeynman}
\end{equation}
This generalizes the Haussmann--Mori--MacDonald relation, Eq.~\eqref{eq:GM}, to interactions of arbitrary many-body rank. 
One should therefore understand the Hausmann--Mori-Macdonald formula as a generalized sum rule 

\begin{equation}
\int_{-\infty}^{\mu_-}\!\mathrm dE\,E\,\rho^-_E  
= \lim_{\lambda \to 1} \langle\Psi_0^\lambda|\partial_\lambda[\lambda {\mathcal H}_\lambda]|\Psi_0^\lambda\rangle.
\label{eq:spectralHellmannFeynman}
\end{equation}
for any interaction rank. This identity requires only small deviation from $\lambda = 1$. It reduces to Eq (1) for pairwise interactions in a flat band.

Integrating now Eq.~\eqref{eq:energyFlow}, since $\lambda E_{\text{G.S.}}(\lambda)$ does not contribute at $\lambda =0$, gives the ground-state energy of the original physical Hamiltonian, $E_{\text{G.S.}}\equiv E_{\text{G.S.}}(\lambda=1)$, 

\begin{equation}
E_{\text{G.S.}}=\int_0^1\!\mathrm d\lambda
\int_{-\infty}^{+\infty}\!\frac{\mathrm d\omega}{2\pi i}\,
e^{i\omega0^+}\omega\operatorname{Tr}[\mathcal G_\lambda(\omega)].
\label{eq:main}
\end{equation}
This principal result gives an exact, nonperturbative relation between the ground-state energy and the interacting propagators. Here $\mathcal G_\lambda$ is the exact propagator for Eq.~\eqref{eq:Hlambda}, with the fixed one-body term and all interactions retained. When only pairwise interactions are present, Eq.~\eqref{eq:main} is consistent with Eq.~\eqref{eq:GMfull} and, in the flat-band limit, reduces to Eq.~\eqref{eq:GM}.

An equivalent form of Eq.~\eqref{eq:main} follows from rescaling fermionic fields $\tilde c^{}=\sqrt\lambda c^{}$, for which $\mathcal H[\tilde c^{\dag},\tilde c^{}] = \lambda \mathcal H_\lambda[c^{\dag},c^{}]$. At fixed coupling, this changes only the field normalization in the same Hamiltonian; $\{\tilde c^{}_{\alpha},\tilde c^{\dag}_{\beta}\}=\lambda\delta_{\alpha\beta}$ and $\widetilde{\mathcal G}_\lambda=\lambda\mathcal G_\lambda$ then require the weight $\mathrm d\lambda/\lambda$ for the rescaled propagator. For the more general choice $\tilde c^{}=f(\lambda) c^{}$, the Hamiltonian $\mathcal H_\lambda=f^{-2}\mathcal H[\tilde c^{\dag},\tilde c^{}]$ has couplings $f^{2n-2}$, with weights $2ff'$ for the canonical propagator and $2f'/f$ for its rescaled counterpart. The ground state energy is independent of the choice for a smooth path $f\in C([0,1])\cap C^1((0,1])$ that is positive for $0<\lambda\le1$,  threading from $f(0)=0$ to $f(1)=1$.

\

\textit{Discussion.---} We have obtained an exact, nonperturbative expression for the ground-state energy in terms of the exact interacting propagators $\mathcal G_{\lambda}$, for arbitrary many-body interactions. The construction applies to instantaneous, number-conserving, normal-ordered interactions, with no restriction to interaction types. Fractional Chern insulators provide the motivation, while the same construction \eqref{eq:main} applies to other correlated fermionic phases with genuine or effective many-body interaction terms.

The auxiliary Hamiltonian path retains residual dispersion through the exact propagators in Eq.~\eqref{eq:main}. Relation~\eqref{eq:main} applies directly to fractional Chern insulators, including those with finite band dispersion, and competing quantum phases in twisted transition metal dichalcogenides.

\

\textit{Acknowledgments}.  This research was partially supported (F.D.M.H.)  by NSF through the Princeton University (PCCM) Materials Research Science and Engineering Center DMR-2614599,  and  (A.K.) from SNSF Grant No. 225225 and  support received from the Simons Foundation (Award No. 566116, P.I. Shinsei Ryu).  A.K. acknowledges additional support under DFG Heisenberg Programme commencing in 2027.

\bibliography{refs}

@article{GalitskiiMigdal1958,
  author = {Galitskii, V. M. and Migdal, A. B.},
  title = {Application of quantum field theory methods to the many body problem},
  journal = {Zh. Eksp. Teor. Fiz.}, volume = {34}, pages = {139}, year = {1958},
  note = {[Sov. Phys. JETP \textbf{7}, 96 (1958)]},
  url = {http://jetp.ras.ru/cgi-bin/dn/e_007_01_0096.pdf}
}

@book{Mahan,
  author = {Mahan, G. D.},
  title = {Many-Particle Physics},
  edition = {3rd}, publisher = {Kluwer Academic/Plenum}, address = {New York}, year = {2000}
}

@article{Haussmann1996,
  author = {Haussmann, R. and Mori, H. and MacDonald, A. H.},
  title = {Correlation Energy and Tunneling Density of States in the Fractional Quantum {H}all Regime},
  journal = {Phys. Rev. Lett.}, volume = {76}, pages = {979}, year = {1996}, doi = {10.1103/PhysRevLett.76.979}
}

@article{MacDonald1999,
  author = {MacDonald, A. H.},
  title = {Correlations weak and strong: divers guises of the two-dimensional electron gas},
  journal = {Int. J. Mod. Phys. B}, volume = {13}, pages = {447}, year = {1999},
  doi = {10.1142/S0217979299000345}, arxivid = {cond-mat/9801037}
}

@article{SodemannMacDonald2013,
  author = {Sodemann, I. and MacDonald, A. H.},
  title = {Landau level mixing and the fractional quantum {H}all effect},
  journal = {Phys. Rev. B}, volume = {87}, pages = {245425}, year = {2013}, doi = {10.1103/PhysRevB.87.245425},
  arxivid = {1302.3896}
}

@article{ZhangSong2024,
  author = {Zhang, L. and Song, X.-Y.},
  title = {{M}oore-{R}ead state in half-filled moir\'e {C}hern band from three-body pseudopotential},
  journal = {Phys. Rev. B}, volume = {109}, pages = {245128}, year = {2024},
  doi = {10.1103/PhysRevB.109.245128}, arxivid = {2403.11478}
}

@article{MooreRead1991,
  author = {Moore, G. and Read, N.},
  title = {Nonabelions in the fractional quantum {H}all effect},
  journal = {Nucl. Phys. B}, volume = {360}, pages = {362}, year = {1991},
  doi = {10.1016/0550-3213(91)90407-O}
}

@article{GreiterWenWilczek1992,
  author = {Greiter, M. and Wen, X.-G. and Wilczek, F.},
  title = {Paired {H}all states},
  journal = {Nucl. Phys. B}, volume = {374}, pages = {567}, year = {1992},
  doi = {10.1016/0550-3213(92)90401-V}
}

@article{ReadRezayi1999,
  author = {Read, N. and Rezayi, E.},
  title = {Beyond paired quantum {H}all states: Parafermions and incompressible states in the first excited {L}andau level},
  journal = {Phys. Rev. B}, volume = {59}, pages = {8084}, year = {1999},
  doi = {10.1103/PhysRevB.59.8084}, arxivid = {cond-mat/9809384}
}

@article{Carbone2013,
  author = {Carbone, A. and Cipollone, A. and Barbieri, C. and Rios, A. and Polls, A.},
  title = {Self-consistent {G}reen's functions formalism with three-body interactions},
  journal = {Phys. Rev. C}, volume = {88}, pages = {054326}, year = {2013},
  doi = {10.1103/PhysRevC.88.054326}, arxivid = {1310.3688}
}

@book{Hellmann1937,
  author = {Hellmann, H.},
  title = {Kvantovaya khimiya},
  publisher = {Glavnaya Redaktsiya Tekhniko-Teoreticheskoi Literatury},
  address = {Moscow and Leningrad}, year = {1937},
  note = {[in Russian]; abridged and revised {German} edition: \emph{Einf{\"u}hrung in die Quantenchemie} (Franz Deuticke, Leipzig und Wien, 1937)}
}

@article{Feynman1939,
  author = {Feynman, R. P.},
  title = {Forces in Molecules},
  journal = {Phys. Rev.}, volume = {56}, pages = {340}, year = {1939}, doi = {10.1103/PhysRev.56.340}
}

@article{Koltun1972,
  author = {Koltun, D. S.},
  title = {Total Binding Energies of Nuclei, and Particle-Removal Experiments},
  journal = {Phys. Rev. Lett.}, volume = {28}, pages = {182}, year = {1972},
  doi = {10.1103/PhysRevLett.28.182}
}

@article{Haldane1983,
  author = {Haldane, F. D. M.},
  title = {Fractional Quantization of the {Hall} Effect: A Hierarchy of Incompressible Quantum Fluid States},
  journal = {Phys. Rev. Lett.}, volume = {51}, pages = {605}, year = {1983},
  doi = {10.1103/PhysRevLett.51.605}
}

@article{SimonRezayiCooper2007,
  author = {Simon, S. H. and Rezayi, E. H. and Cooper, N. R.},
  title = {Pseudopotentials for multiparticle interactions in the quantum {Hall} regime},
  journal = {Phys. Rev. B}, volume = {75}, pages = {195306}, year = {2007},
  doi = {10.1103/PhysRevB.75.195306}
}
\end{document}